\documentclass[12pt,english,american,prl,floatfix,superscriptaddress,twocolumn,showpacs,amsmath,amssymb,reprint,sort]{revtex4-1}
\usepackage[T1]{fontenc}
\usepackage[latin9]{inputenc}
\usepackage{amsmath}
\usepackage{amssymb}
\usepackage{graphicx}
\usepackage{esint}

\makeatletter
\PassOptionsToPackage{caption=false}{subfig}
\usepackage{hyperref}
\hypersetup{
breaklinks=true,
colorlinks=true,
citecolor=blue,
linkcolor=blue,
filecolor=blue,
urlcolor=blue
}
\IfFileExists{lmodern.sty}{\usepackage{lmodern}}{}

\makeatother

\usepackage{babel}
\begin{document}
\selectlanguage{english}%

\title{Prediction of Nonlinear Evolution Character of Energetic-Particle-Driven
Instabilities}

\author{V. N. Duarte}
\email{vnduarte@if.usp.br, vduarte@pppl.gov}

\selectlanguage{english}%

\address{Institute of Physics, University of São Paulo, São Paulo, SP, 05508-090,
Brazil}

\address{Princeton Plasma Physics Laboratory, Princeton University, Princeton,
NJ, 08543, USA}

\author{H. L. Berk}

\address{Institute for Fusion Studies, University of Texas, Austin, TX, 78712,
USA}

\author{N. N. Gorelenkov}

\address{Princeton Plasma Physics Laboratory, Princeton University, Princeton,
NJ, 08543, USA}

\author{W. W. Heidbrink}

\address{University of California, Irvine, CA, 92697, USA}

\author{G. J. Kramer}

\address{Princeton Plasma Physics Laboratory, Princeton University, Princeton,
NJ, 08543, USA}
\selectlanguage{american}%

\author{R. Nazikian}
\selectlanguage{english}%

\address{Princeton Plasma Physics Laboratory, Princeton University, Princeton,
NJ, 08543, USA}

\author{D. C. Pace}

\address{General Atomics, San Diego, CA, 92186, USA}

\author{M. Podestà}

\address{Princeton Plasma Physics Laboratory, Princeton University, Princeton,
NJ, 08543, USA}

\author{B. J. Tobias}

\address{Princeton Plasma Physics Laboratory, Princeton University, Princeton,
NJ, 08543, USA}

\author{M. A. Van Zeeland}

\address{General Atomics, San Diego, CA, 92186, USA}

\date{\today}
\selectlanguage{american}%
\begin{abstract}
A general criterion is proposed and found to successfully predict
the emergence of chirping oscillations of unstable Alfvénic eigenmodes
in tokamak plasma experiments. The model includes realistic eigenfunction
structure, detailed phase-space dependences of the instability drive,
stochastic scattering and the Coulomb drag. The stochastic scattering
combines the effects of collisional pitch angle scattering and micro-turbulence
spatial diffusion. The latter mechanism is essential to accurately
identify the transition between the fixed-frequency mode behavior
and rapid chirping in tokamaks and to resolve the disparity with respect
to chirping observation in spherical and conventional tokamaks.
\end{abstract}
\maketitle
\selectlanguage{english}%
In fusion grade plasmas, there is a population of energetic particles
(EPs) with typical energies substantially greater than those of the
thermal background. These particles provide an energy-inverted population,
that through the kinetic wave-particle interaction with Alfvén waves,
can induce instabilities that jeopardize plasma confinement \citep{Heidbrink2008,Gorelenkov2014}.
The nature of these oscillations vary considerably (with the possibility
of several bifurcations \citep{BerkPRL1996}), with two typical non-linear
scenarios being: (a) the excitation of a slow evolving amplitude,
nearly fixed-frequency oscillation and (b) coherent oscillations that
chirp in frequency at timescales much shorter than that of the plasma
equilibrium modification. These scenarios lead to dominant diffusive
and convective transport of EPs, respectively. 

\selectlanguage{american}%
This letter addresses two outstanding and inter-connected issues \foreignlanguage{english}{that,
in spite of their major relevance for the transport of EPs in future-generation
burning plasmas, are currently not understood}. The first issue is
\foreignlanguage{english}{what plasma conditions most strongly determine
the }likelihood of each non-linear saturation scenario in experiments.
\foreignlanguage{english}{The second is why the chirping response
(observed in all major tokamaks, e.g. DIII-D \citep{Heidbrink1995Chirping},
NSTX \citep{FredricksonPoP2006,Podest2012}, JET \citep{Boswell2006},
MAST \citep{Pinches2004}, JT-60U \citep{Kusama1999,Kramer2000},
ASDEX-U \citep{Horvarth2016Chirping}) is much more common in spherical
tokamaks than in conventional tokamaks. This classification is important
in anticipating whether EP-induced instabilities in burning plasma
experiments will likely lead to steady oscillations, where quasi-linear
theory \citep{VedenovSagdeev1961,Drummond_Pines_1962,Berk1995LBQ}
would be expected to described EP transport or chirping, which would
then require new theoretical tools to assess the consequences of the
induced EP transport. }

In this letter we show that a previous approach that attempted to
simplify the needed input that the theory requires \citep{Lilley2009PRL}
is insightful but limited in making accurate predictions for experimental
scenarios. Here we employ a generalized formulation and show that
its predictions are in accordance with observations. This analysis
reveals that micro-turbulence, even while producing no observable
effect on beam ion transport, provides the vital mechanism in determining
which non-linear regime is more likely for a mode as well as the mode
transition from one regime to the other, as parameters of an experiment
change in time.

\selectlanguage{english}%
We focus the analysis on the onset of a mode non-linear evolution
near marginal stability. The interaction Hamiltonian between a particle
(at position $\mathbf{r}$, with velocity $\mathbf{v}$) and a tokamak
eigenmode with frequency $\omega$ can be written as $q\mathbf{A}\left(\mathbf{r},t\right)\cdotp\mathbf{v}=C\left(t\right)\underset{j}{\sum}V_{j}\left(\mathcal{E},P_{\varphi},\mu\right)e^{i\left(j\theta_{a}-n\varphi_{a}-\omega t\right)}$,
where $\mathbf{A}\left(\mathbf{r},t\right)$ is the perturbed vector
potential along an unperturbed orbit in a gauge where the electrostatic
potential vanishes, $\mathcal{E}$ is the unperturbed energy, $P_{\varphi}$
the canonical angular momentum, $\mu$ the magnetic moment (all per
unit EP mass), the summation is over all integers $j$, $\varphi_{a}$
and $\theta_{a}$ are the action angles in the toroidal and poloidal
directions, $q$ is the charge of an EP and $n$ is a fixed quantum
number for the angular response of a perturbed linear wave in an axisymmetric
toroidal tokamak. $V_{j}$ (see Eq. 12 of Ref. \citep{BreizmanPoP1997})
accounts for the wave-particle energy exchange. Upon a suitable normalization,
the amplitude $C(t)$ has been shown to be governed by an integro-differential
cubic equation that is nonlocal in time \citep{BerkPRL1996,Lilley2009PRL,BerkPPR1997}\footnote{The cubic equation was derived independently for vortex flow in fluids;
see \citep{Hickernell1984}}, 
\begin{equation}
\begin{array}{c}
\frac{dC(t)}{dt}-C(t)=-\underset{j}{\sum}\intop d\Gamma\mathcal{H}\int_{0}^{t/2}d\tau\tau^{2}C\left(t-\tau\right)\times\\
\times\int_{0}^{t-2\tau}d\tau_{1}e^{-\hat{\nu}_{stoch}^{3}\tau^{2}\left(2\tau/3+\tau_{1}\right)+i\hat{\nu}_{drag}^{2}\tau\left(\tau+\tau_{1}\right)}\times\\
\times C\left(t-\tau-\tau_{1}\right)C^{*}\left(t-2\tau-\tau_{1}\right)
\end{array}\label{eq:cubic-1}
\end{equation}
where $\mathcal{H}=2\pi\omega\delta\left(\omega-\Omega_{j}\right)\left|V_{j}\right|^{4}\left(\frac{\partial\Omega_{j}}{\partial I}\right)^{3}\frac{\partial f}{\partial\Omega}$,
with $f$ being the equilibrium distribution function. We assume a
low frequency mode for which $\mu$ is conserved. Then $\partial/\partial I\equiv-n\partial/\partial P_{\varphi}+\omega\partial/\partial\mathcal{E}$.
The resonance condition is given by $\Omega_{j}=\omega+n\omega_{\varphi}-j\omega_{\theta}\approx0$,
where $\omega_{\theta}$ and $\omega_{\varphi}$ are the mean poloidal
and toroidal transit frequencies of the equilibrium orbit. The phase-space
integration is given by$\intop d\Gamma...=\left(2\pi\right)^{3}\underset{\sigma_{\parallel}}{\sum}\int dP_{\varphi}\int d\mathcal{E}/\omega_{\theta}\int m_{EP}cd\mu/q...$,
where $m_{EP}$ is the mass of EPs, $c$ is the light speed and $\sigma_{\parallel}$
accounts for counter- and co-passing particles. The effective collisional
operator can be cast in the form $C[f]=\nu_{scatt}^{3}\frac{\partial^{2}f}{\partial\Omega^{2}}+\nu_{drag}^{2}\frac{\partial f}{\partial\Omega}$,
where $\nu_{scatt}$ and $\nu_{drag}$ are understood to be the effective
pitch-angle scattering and drag (slowing down) coefficients, defined
in Eq. 6 of Ref. \citep{Lilley2009PRL}. $\nu_{stoch}$ is the effective
stochasticity, which includes $\nu_{scatt}$. In equation \eqref{eq:cubic-1},
the circumflex denotes normalization with respect to $\gamma=\gamma_{L}-\gamma_{d}$
(growth rate minus damping rate) and $t$ is the time normalized with
the same quantity. Vlasov simulation codes have shown \citep{BerkPLA1997,BerkCandy1999}
that the blow-up solutions of \eqref{eq:cubic-1} are precursors to
chirping behavior. 

\begin{figure}
\begin{centering}
\includegraphics[scale=0.3]{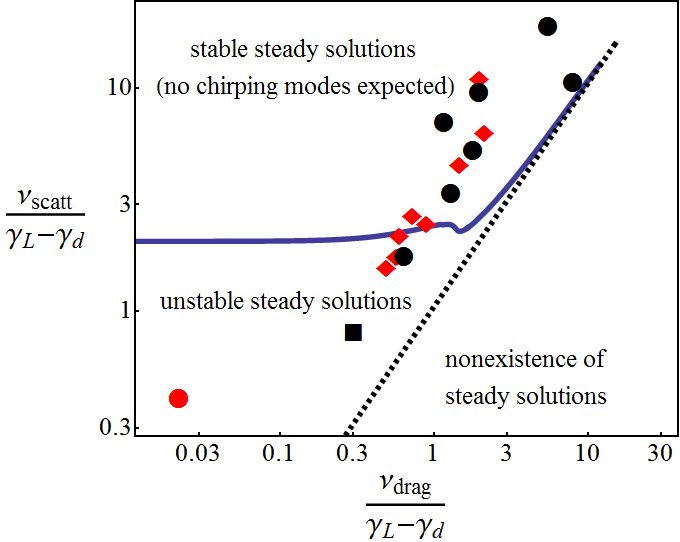}
\par\end{centering}
\caption{Comparison between analytical predictions with experiment when single
characteristic values for phase space parameters are chosen. The dotted
line delineates the region of existence of steady amplitude solutions
of the cubic equation \eqref{eq:cubic-1} while the solid line delineates
the region of stability, as predicted by \citep{Lilley2009PRL}. Modes
that chirped are represented in red and the ones that were steady
are in black, as experimentally observed in DIII-D (disks), NSTX (diamonds)
and TFTR (square).\label{fig:Comparison-Lilley}}
\end{figure}

The type of nonlinear evolution of a wave destabilized by a perturbing
EP drive is strongly dependent on the kernel of the integrals of Eq.
\eqref{eq:cubic-1}, more specifically on the ratio between the effective
stochastic relaxation felt by the EPs and the effective drag rate,
as well as the linear growth rate. In Ref. \citep{Lilley2009PRL},
Eq. \eqref{eq:cubic-1} was simplified \foreignlanguage{american}{by
using characteristic values for }the collisional $\nu_{scatt}$ and
$\nu_{drag}$ and conditions for the existence and stability of solutions
of the cubic equation were derived. In Fig. \ref{fig:Comparison-Lilley},
we test for the first time this prediction against modes measured
in different tokamaks. In order to determine mode properties, we employ
the kinetic-MHD code NOVA \citep{Gorelenkov1999ChengFu} to compute
eigenstructures and the frequency continua and gaps. Its kinetic postprocessor
NOVA-K \citep{CHENG1992,Gorelenkov1999Saturation} is used to calculate
perturbative contributions that can stabilize and destabilize MHD
eigenmodes. In addition, NOVA-K is also employed to compute resonant
surfaces in $\left(\mathcal{E},P_{\varphi},\mu\right)$ space. In
order to characterize the mode being observed in the experiment, NSTX
reflectometer measurements are compared to the mode structures computed
by NOVA, employing a similar procedure as the one used in Refs. \citep{Fredrickson2009,Podest2012}.
In DIII-D, similar identification is performed using Electron Cyclotron
Emission (ECE) \citep{VanZeeland2006}. 

\selectlanguage{american}%
W\foreignlanguage{english}{e see from Fig. \ref{fig:Comparison-Lilley}
that about half of the chirping NSTX modes lie in a region where a
stable steady mode is predicted by Ref. \citep{Lilley2009PRL}. For
the DIII-D experimental cases that produced fixed-frequency modes,
the predictions of Ref. \citep{Lilley2009PRL} are mostly in agreement
although one point is borderline and another one may be unstable enough
to be in a chirping regime. Hence we see that using the simplified,
although elaborate, modeling akin to that used in \citep{Lilley2009PRL},
might be in satisfactory agreement with DIII-D data but is generally
not satisfactory for much of the NSTX and TFTR data. This comparison
indicates that the use of a single characteristic value, being representative
of the entire phase space,} for \foreignlanguage{english}{$\nu_{scatt}$
(considered the only contribution to $\nu_{stoch}$) and $\nu_{drag}$,
although very insightful, appears insufficient to provide quantitative
predictions for practical tokamak cases. This conclusion motivated
the pursuit of a general theoretical prediction to take into account
important missing elements, such as spatial mode structures and local
phase-space contributions on multiple resonant surfaces of the wave-particle
interaction terms, all of which are needed in toroidal geometry. The
appropriate weightings for the various needed quantities can be expressed
in the action-angle formulation. A necessary, although not sufficient,
condition for chirping solutions is that the right hand side of \eqref{eq:cubic-1}
be positive. The resonance condition, $\delta\left(\omega-\Omega_{l}\left(P_{\varphi},\mathcal{E},\mu\right)\right)$,
allows one of the phase-space integrals to be eliminated. Upon integration
over $\tau_{1}$ and redefinition of the integration variable $z=\nu_{drag}\tau$
one finds the following criterion for the non-existence of steady
solutions of \eqref{eq:cubic-1}:}

\selectlanguage{english}%
\begin{equation}
Crt=\frac{1}{N}\underset{j,\sigma_{\parallel}}{\sum}\int dP_{\varphi}\int d\mu\frac{\left|V_{j}\right|^{4}}{\omega_{\theta}\nu_{drag}^{4}}\left|\frac{\partial\Omega_{j}}{\partial I}\right|\frac{\partial f}{\partial I}Int<0\label{eq:Criterion-1}
\end{equation}
where
\begin{equation}
Int\equiv Re\int_{0}^{\infty}dz\frac{z}{\frac{\nu_{stoch}^{3}}{\nu_{drag}^{3}}z-i}exp\left[-\frac{2}{3}\frac{\nu_{stoch}^{3}}{\nu_{drag}^{3}}z^{3}+iz^{2}\right]\label{eq:Int-1}
\end{equation}

\begin{figure}
\begin{centering}
\includegraphics[scale=0.3]{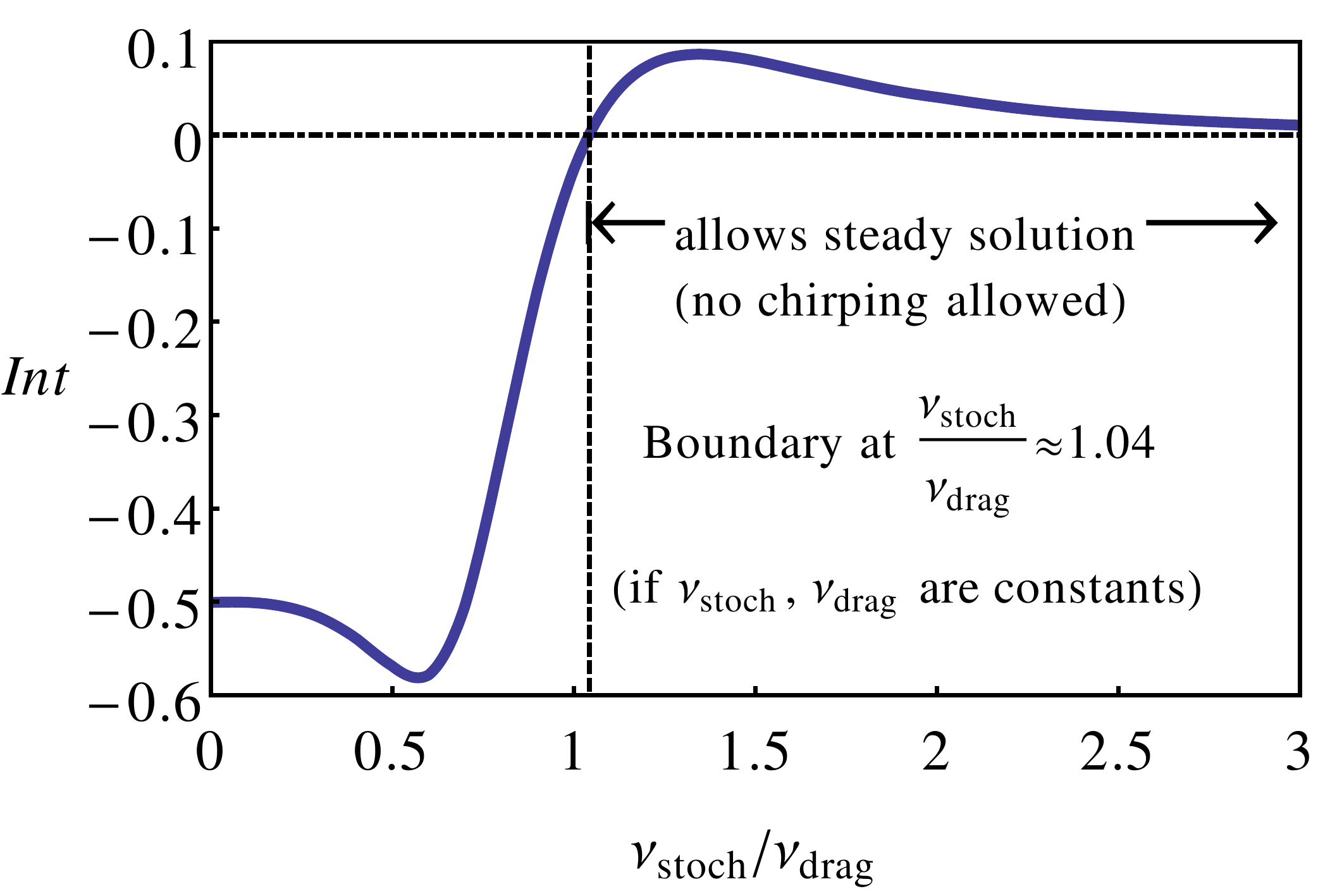}
\par\end{centering}
\caption{$Int$ (Eq. \eqref{eq:Int-1}) plotted in terms of local values of
$\nu_{stoch}/\nu_{drag}$.\label{fig:Int}}
\end{figure}
For the resonances to be linearly destabilizing to positive energy
waves, \textit{Int }(plotted in Fig. \ref{fig:Int}) is the only component
of the criterion \eqref{eq:Criterion-1} that can be negative from
the phase-space regions which contribute positively to the instability
growth. $N$ is a normalization factor consisting of the same sum
that appears in Eq. \eqref{eq:Criterion-1} except for $Int$. Thus
we use the contributions from each resonance weighted in accord with
the appropriate eigenfunction (that fits the measured field structure)
and the position in phase space of the resonant interaction. We will
see that this procedure produces quite a different conclusion from
the less detailed method that uses a single characteristic factor,
as is the case in Fig. \ref{fig:Comparison-Lilley}.

Non-steady oscillations, with the likelihood of chirping, are predicted
to occur if $Crt<0$ while a steady (fixed-frequency) solution exists
if $Crt>0$. However, we see from Fig. \ref{fig:Int} that the magnitude
of $Int$ could be an order of magnitude larger in phase space regions
where $\nu_{stoch}/\nu_{drag}\lesssim1.04$ compared with regions
where $\nu_{stoch}/\nu_{drag}\gtrsim1.04$. Hence, because of this
disparity, it can turn out that a choice of the use of a single characteristic
value for $\nu_{stoch}/\nu_{drag}$, would lead to a positive value
for $Crt$ while the use of the appropriately weighted average leads
to a negative value for $Crt$. Such a change is indeed the case for
all the TFTR and DIII-D modes and for most of the NSTX modes shown
in Fig. \ref{fig:Comparison-Lilley}, where $\nu_{stoch}$ was considered
simply as $\nu_{scatt}$. The reason for this sensitivity is that
there will always be a contribution to $Crt$ from a phase space region
where $\nu_{scatt}/\nu_{drag}\ll1$ because the pitch angle scattering
coefficient goes to zero as $\mu$ vanishes. Hence even when the characteristic
value of $\nu_{scatt}/\nu_{drag}$ is substantially greater than unity,
one still can find that $Crt<0$. 

The above observation indicates that pitch-angle scattering $\nu_{scatt}$
may not always be the dominant mechanism in determining $\nu_{stoch}$.
Hence, we now introduce the contribution of fast-ion electrostatic
micro-turbulence for the determination of $\nu_{stoch}$ with the
following procedure. The TRANSP code \citep{Hawryluk1980} is employed
to obtain the thermal ion radial thermal conductivity, $\chi_{i}$
(which is essentially the particle diffusivity, $D_{i}$ \citep{Heidbrink2009PRL})
based on power balance. The heat diffusivity due to collisions is
subtracted out and the remaining diffusivity is attributed to micro-turbulence
interaction with the ions. Then the EPs diffusivity is estimated by
using the scalings determined in a gyrokinetic simulation \citep{ZhangLinChen2008PRL},
which for passing particles gives $D_{EP}\approx5D_{i}T_{i}/E_{EP}$.
In the experiments we analyzed, the drive was mostly from the passing
particles and therefore we used this relation as an estimate for $D_{EP}$.
The response of the resonant EPs to perturbing fields is essentially
one-dimensional \citep{BerkPPR1997} and produces steep gradients
in the EP distribution in this perturbing direction. We can then accurately
account for the diffusion that is directed in all phase space directions,
by projecting the actual diffusion from all these directions onto
the steepest gradient path defined by the one-dimensional dynamics,
using the specific relation given by Eq. (2) of Ref. \citep{LangFu2011}. 

\begin{figure}
\begin{centering}
\includegraphics[scale=0.19]{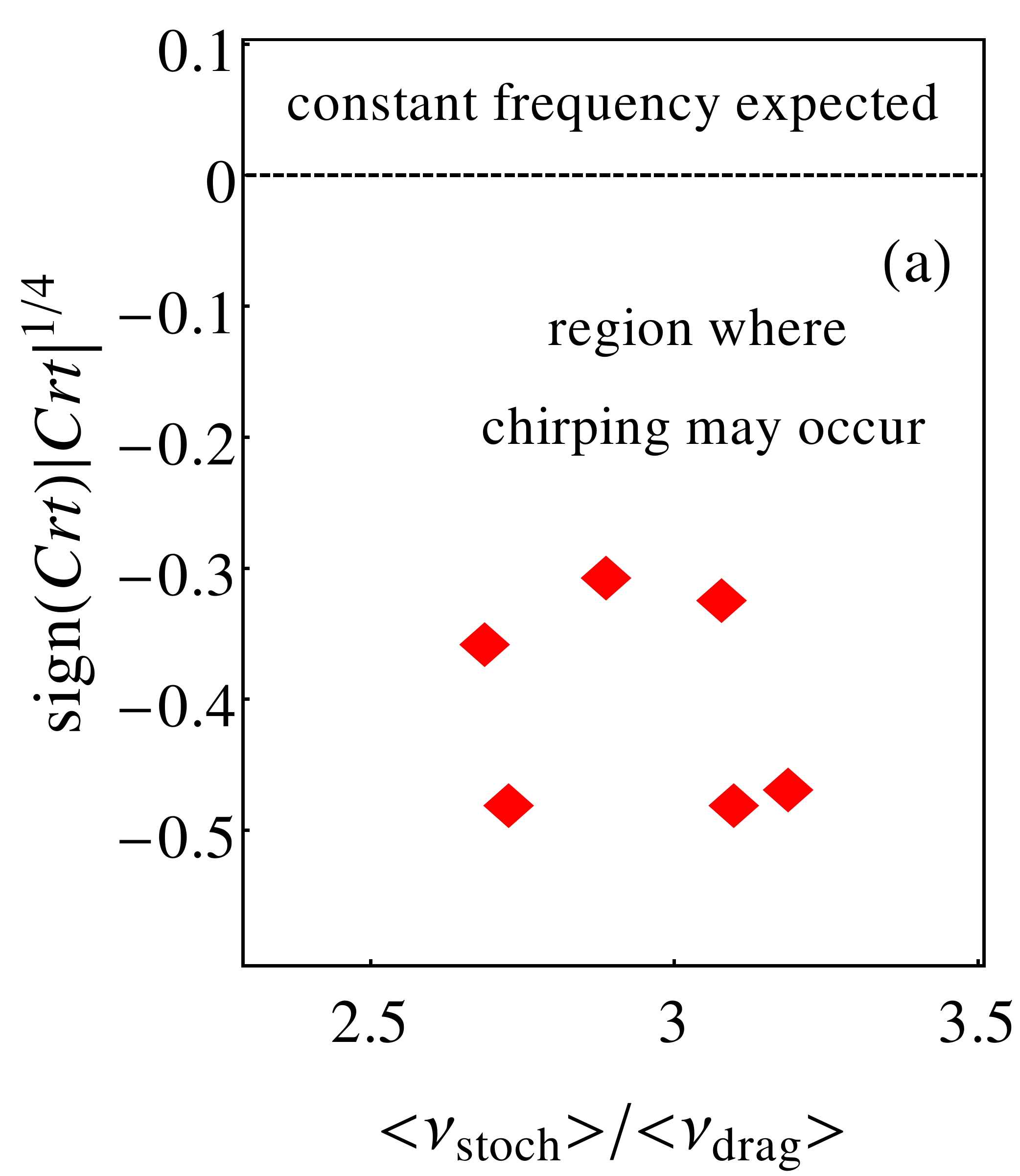}\includegraphics[scale=0.2]{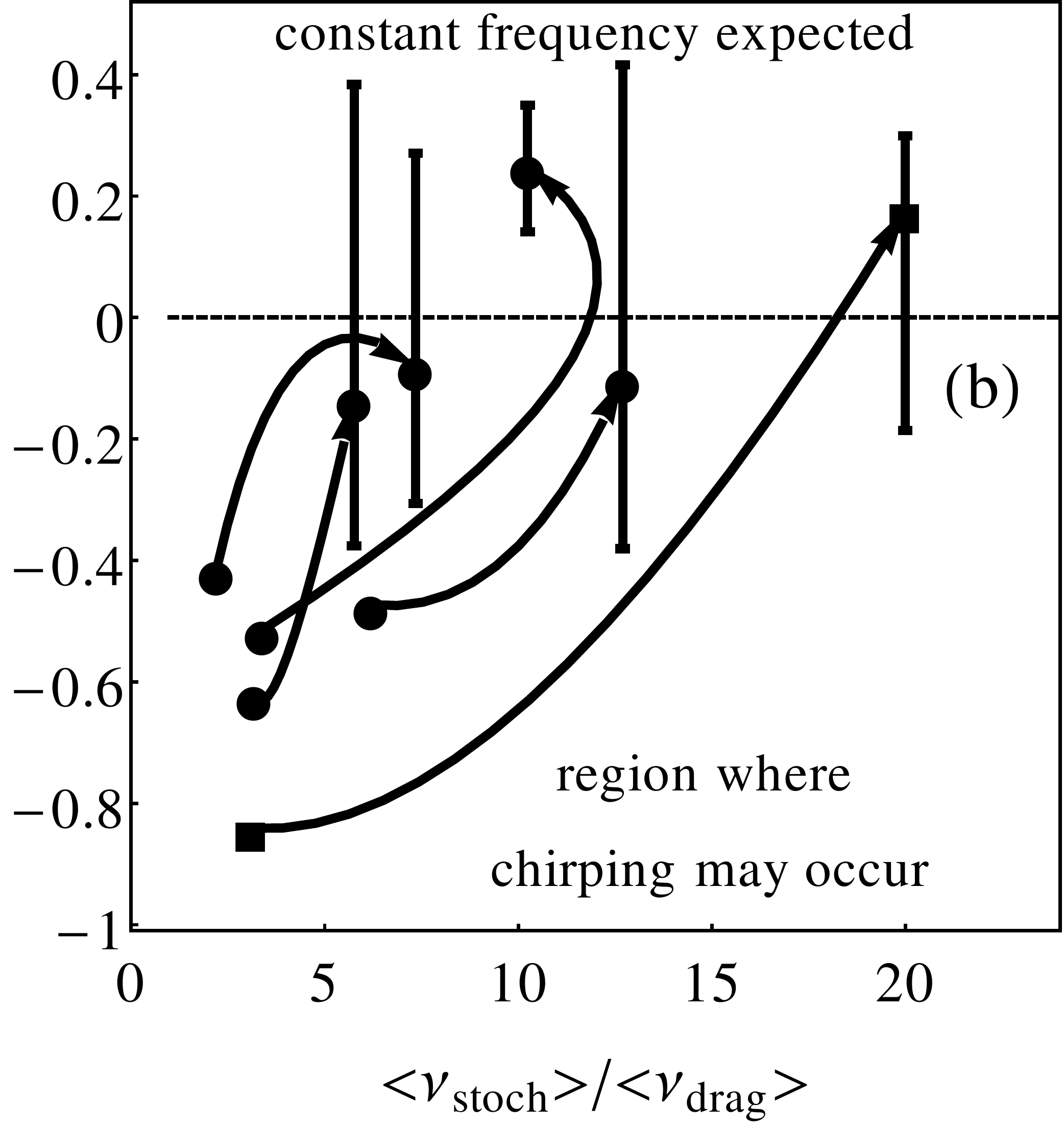}
\par\end{centering}
\caption{Numerical values for $|Crt|^{1/4}$ multiplyed by the sign of $Crt$
as a function of $<\nu_{stoch}>/<\nu_{drag}>$. Modes that chirped
are represented in red and the ones that were steady are in black,
as experimentally observed in (a) NSTX (diamonds) and (b) DIII-D (disks)
and TFTR (squares). The arrows represent the effect of micro-turbulence.
The bars represent by how much the prediction for the modes change
if we double the turbulent diffusity (upper bars) or divide it by
2 (lower bars). In NSTX case, the points hardly move upon the addition
of spatial diffusion to collisional scattering. \label{fig:Effect-of-Turb}}
\end{figure}

Fig. \ref{fig:Effect-of-Turb} shows values of $|Crt|^{1/4}$ multiplyed
by the sign of $Crt$, as a function of the ratio of phase-space averaged
stochasticity and drag for modes of Fig. \ref{fig:Comparison-Lilley}.
This representation provides better visualization than simply plotting
$Crt$, especially close to the steady/chirping boundary and is chosen
because of the fourth power dependence $Int\approx1.022\left(\nu_{stoch}/\nu_{drag}\right)^{-4}$
for $\nu_{stoch}/\nu_{drag}\gg1$. Fig. \ref{fig:Effect-of-Turb}(a)
shows chirping modes in NSTX and Fig. \ref{fig:Effect-of-Turb}(b)
shows steady modes in DIII-D and TFTR. The curved arrows represent
how the prediction for a mode is affected by micro-turbulence-induced
scattering of EPs. It has a strong effect on DIII-D and TFTR (bringing
the modes to the steady region, or at least very close to it) while
its effect is imperceptible for the chirping modes in NSTX. This is
because, unlike in conventional tokamaks, thermal ion transport in
spherical tokamaks (STs) is usually very close to neoclassical levels
\citep{KayeNF2007,Fields2004} even though the electron transport
is anomalous. NSTX modes in Fig. \ref{fig:Effect-of-Turb}(a) are
only able to transition to the fixed-frequency region when $\nu_{stoch}$
is artificially multiplied by a factor from 10 to 50, depending on
the specific mode, which indicates the robustness of the chirping
prediction.

\begin{widetext}

\begin{figure}
\selectlanguage{american}%
\begin{centering}
\includegraphics[scale=0.42]{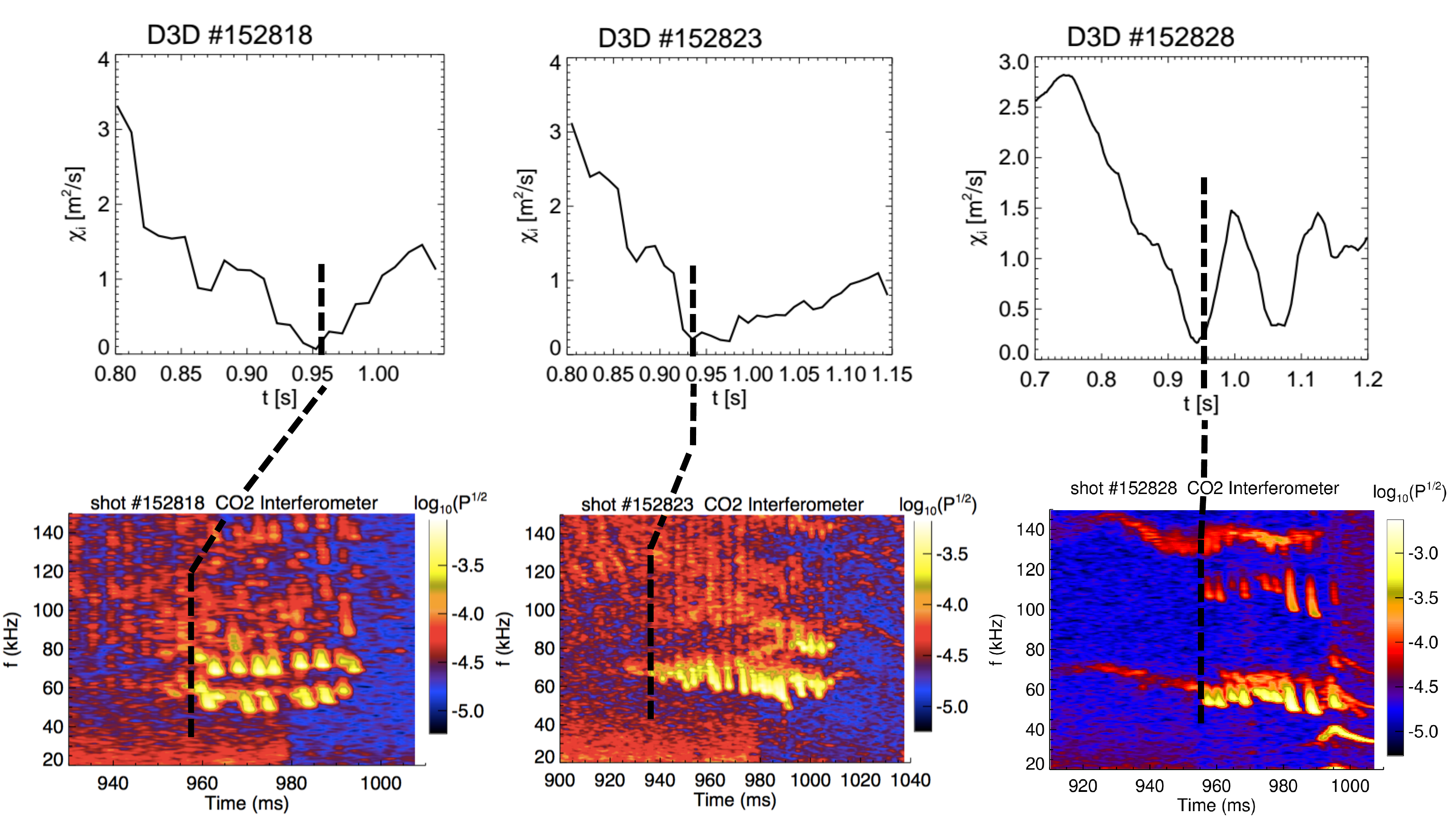}
\par\end{centering}
\selectlanguage{english}%
\caption{Correlation in DIII-D between the emergence of chirping and the development
of low diffusivity, as calculated by TRANSP at the radius where the
mode is peaked. \label{fig:Correlation-in-DIII-D}}
\end{figure}

\end{widetext}

Guided by this theory, we have then examined chirping modes that rarely
appear in DIII-D tokamak and we have found that chirping onset in
DIII-D is observed to correlate very closely with conditions where
thermal ion transport had drastically decreased, as shown in Fig.
\ref{fig:Correlation-in-DIII-D}. This is attributed to the decrease
in micro-turbulence-induced transport, which also causes decreased
EP transport. Alfvénic modes only started chirping when the thermal
ion conductivity dropped to values lower than $0.3m^{2}/s$. An example
of the evaluation of the criterion \eqref{eq:Criterion-1} is DIII-D
shot 152828 (Fig. \ref{fig:Correlation-in-DIII-D} (c)). Before chirping
starts (at $t=920ms$, when $D_{th,i}\approx0.55m^{2}/s$) the calculated
criterion is $Crt=+0.001$. During the early phase of chirping (at
$t=955ms$, when $D_{th,i}\approx0.25m^{2}/s$) the value is $Crt=-0.013$,
i.e. the mode has transitioned from the positive (steady) region to
the negative region of $Crt$, therefore allowing chirping, in agreement
with the observation. 

The micro-turbulence interaction with EPs is a key factor that determines
the nature of mode saturation regime (quasi-steady and chirping) and
also the transition between them. It also explains the longstanding
question of why chirping Alfvénic modes are ubiquitous in STs and
rare in conventional tokamaks. \foreignlanguage{american}{Experimentally,
the EP transport due to micro-turbulence is too low compared to Alfvénic-induced
transport \citep{Pace2013,Geiger2015PPCF}, yet the turbulent interaction
remains capable of drastically changing the non-linear stability regime
of the Alfvénic modes at their onset, when mode amplitude is low.
This suggests that micro-turbulence simulations employed to predict
the thermal plasma transport of future burning plasma devices must
also be factored in to considerations of the drive and saturation
of modes driven by EPs.}

This work provides elements for choosing which of the two extreme
scenarios is most likely to be relevant for predicting the character
of the energetic particle transport, based on the sign of $Crt$.
For a negative $Crt$ the physical conditions are established to enable
a nonlinear BGK-like mode \citep{BGK1957} to form, where the frequency
remains locked to a particle resonance frequency as particles trapped
by the wave are convected in phase space which, for the Alvénic instabilities,
primarily causes resonant energetic particles to flow across field
lines. Alternatively, a positive $Crt$ represents the lack of chirping
and indicates that the details of the nonlinear particle transport
might be described by a quasilinear diffusion theory \citep{VedenovSagdeev1961,Berk1995LBQ,Drummond_Pines_1962}.
Therefore, the application of this criterion should be important in
the planning and modeling of scenarios for future fusion plasma experiments.
\begin{acknowledgments}
We acknowledge fruitful discussions with G.-Y. Fu, E. D. Fredrickson,
B. N. Breizman, W. Wang and W. Guttenfelder and the support of R.
M. O. Galvão. This work was supported by the São Paulo Research Foundation
(FAPESP, Brazil) under grants 2012/22830-2 and 2014/03289-4, and by
US Department of Energy (DOE) under contracts DE-AC02-09CH11466 and
DE-FC02-04ER54698. This work was carried out under the auspices of
the University of São Paulo - Princeton University Partnership, project
``\textit{Unveiling Efficient Ways to Relax Energetic Particle Profiles
due to Alfvénic Eigenmodes in Burning Plasmas}''.
\end{acknowledgments}

\bibliographystyle{apsrev4-1}

\selectlanguage{american}%

\end{document}